\documentclass[twocolumn,floatfix,prx,aps,showpacs]{revtex4-1}
\usepackage{graphicx,amsmath,amssymb,color}
\usepackage{nicefrac}
\usepackage{multirow, makecell, ragged2e}
\usepackage[titletoc,title]{appendix}

\newcommand{\be}{\begin{equation}}
\newcommand{\ee}{\end{equation}}

\newcommand{\ba}{\begin{eqnarray}}
\newcommand{\ea}{\end{eqnarray}}

\renewcommand{\vec}[1]{\mbox{\boldmath$#1$}}

\def\beq{\begin{eqnarray}}
\def\eeq{\end{eqnarray}}
\newcommand{\mi}{\mathrm{i}}
\newcommand\eg{\emph{e.g.}~}
\newcommand\ie{\emph{i.e.}~}

\makeatletter
\newcommand*{\rom}[1]{\expandafter\@slowromancap\romannumeral #1@}
\makeatother

\begin{document}

\title{Berry phase of the composite-fermion Fermi sea: Effect of Landau-level mixing}

\author{Songyang Pu$^1$, Mikael Fremling$^2$ and J. K. Jain$^1$}
\affiliation{$^1$Department of Physics, 104 Davey Lab, Pennsylvania State University, University Park,Pennsylvania 16802,USA.
  \\$^2$Department of Theoretical Physics, Maynooth University, Maynooth, Co.Kildare, W23 HW31, Ireland.}

\date{\today}

\begin{abstract} 
We construct explicit lowest-Landau-level wave functions for the composite-fermion Fermi sea and its low energy excitations following a recently developed approach [Pu, Wu and Jain, Phys. Rev. B {\bf 96}, 195302 (2018)]
and demonstrate them to be very accurate representations of the Coulomb eigenstates.
We further ask how the Berry phase associated with a closed loop around the Fermi circle,
predicted to be $\pi$ in a Dirac composite fermion theory satisfying particle-hole symmetry [D. T.
Son, Phys. Rev. X {\bf 5}, 031027 (2015)],
is affected by Landau-level mixing. For this purpose,
we consider a simple model wherein we determine the variational ground state as a function of Landau-level mixing within the space spanned by two basis functions: the lowest-Landau-level projected and the unprojected composite-fermion Fermi sea wave functions.
We evaluate Berry phase for a path around the Fermi circle within this model following a recent prescription,
and find that it rotates rapidly as a function of Landau-level mixing.
We also consider the effect of a particle-hole symmetry-breaking three-body interaction on the Berry phase while confining the Hilbert space to the lowest Landau level.
Our study deepens the connection between the $\pi$ Berry phase and the exact particle-hole symmetry in the lowest Landau level. 
\end{abstract}

\maketitle

\section{Introduction}

At the half-filled lowest Landau level (LL), the system of strongly correlated electrons undergoes a non-perturbative transmutation into a compressible Fermi sea of weakly interacting composite fermions (CFs) \cite{Halperin93,Kalmeyer92}.
The question of how the particle-hole (PH) symmetry of the original electrons confined to the lowest LL (LLL) manifests for composite fermions in zero effective magnetic field has attracted attention in recent years,
primarily inspired by the work of Son~\cite{Son15,Son16,Son18},
who has proposed that a PH symmetric field theory for the CF Fermi sea (CFFS) can be formulated by treating it as a Fermi sea of Dirac composite fermions.
This is to be contrasted with the Chern-Simons field theory of Halperin, Lee and Read \cite{Halperin93},
which assumes a Fermi sea of non-relativistic composite fermions.
Experiments have been suggested to distinguish between these two formulations\cite{Son15,Wang15,Wang16,Levin17,Nguyen17,Nguyen17b},
although they appear to produce consistent predictions for many quantities of interest~\cite{Wang17,Cheung17,Mulligan16,Kumar18}.
The possibility of a spontaneous breaking of the PH symmetry at the half-filled LL has also been considered\cite{Barkeshli15}. 

In parallel, we have a very precise microscopic theory for the CFFS state in terms of an explicit wave function\cite{Rezayi94,Rezayi00,Shao15,Fremling18,Wang17c,Geraedts17c},
constructed by the standard method of composite-fermionizing \cite{Jain89} the Fermi sea wave function of {\em non-relativistic} electrons by vortex attachment.
This wave function is very close to the exact Coulomb ground state for all cases studied so far, and, as a corollary,
also satisfies the PH symmetry to an excellent approximation.
That raises the natural question: How does the microscopic theory dovetail with the debate on the field theoretical description of the CFFS? A direct path from the microscopic wave function to an effective field theory confined to the LLL is at present unavailable,
but one can aim to test certain sharp consequences of the effective field theory within the microscopic approach.
A fundamental prediction of Son's theory is a $\pi$ Berry phase associated with a closed loop around the Fermi circle. A justification, albeit  not a proof,
for the $\pi$ phase was given in Refs.~\cite{Wang15,Balram16b} starting from the microscopic wave functions of composite fermions.
For an explicit calculation of the Berry phase,
it is not immediately clear how to define the Berry phase for the CFFS,
because the overlap integral between two successive points along the Fermi circle vanishes due to momentum conservation.
Fremling {\em et al.}~\cite{Fremling18} circumvented the problem by considering a closed path for a pair of antipodal CF particles,
so that the center-of-mass (CM) momentum is preserved.
Wang {\em et al.}~\cite{Wang17c} and Geraedts {\em et al.}~\cite{Geraedts17c} define the Berry phase through an overlap integral with one of the wave functions appropriately translated in momentum space through a projected density operator.
They find that the Berry phase associated with the path of a composite fermion enclosing the Fermi sea is exactly $\pi$ provided the wave function satisfies the PH symmetry exactly~\cite{Geraedts17c}.
(This was also a necessary assumption in Ref.~\cite{Balram16b}.) The trial wave function satisfies the PH symmetry to a high degree but not exactly,
and therefore the Berry phase for the trial wave function is close but not equal to $\pi$,
but the Berry phase is exactly $\pi$ if the corresponding exact Coulomb eigenstates are used instead.
Support for $\pi$ Berry phase was also offered by the work of Gearedts {\em et al.}\cite{Geraedts16} who demonstrated an absence of $2k_F$ back-scattering for a PH symmetric disorder. 

The objective of this article is two-fold. The first is to generalize the approach of Pu,
Wu and Jain (PWJ) \cite{Pu17} to construct a LLL wave function for the CFFS,
which is given in Eq. (~\ref{CF Fermi sea}).
We further demonstrate that the wave functions for the ground states and low energy excited states provides accurate approximations for the exact Coulomb eigenstates.
An advantage of this wave function is that it is written as a single Slater determinant and can be evaluated for large systems. 

Second, we ask how robust the Berry phase is to LL mixing, which also breaks PH symmetry.
(PH symmetry can meaningfully be defined only within a given LL.) It has been known since the beginning that some degree of LL mixing, which is always present in experiments,
does not cause any correction to the fractional quantization of the Hall resistance.
The strength of LL mixing is conveniently measured by a parameter $\kappa$, defined as
\be
\kappa\equiv{e^2/\epsilon l\over \hbar\omega_c},
\ee
where $l=\sqrt{\hbar c/eB}$ is the magnetic length, $\epsilon$ is the dielectric constant of the background material,
$\hbar\omega_c=\hbar eB/m_b c$ is the cyclotron energy of electron, and $m_b$ is the electron band mass.
Experiments in low density p-doped GaAs samples \cite{Santos92,Santos92b,Pan05,Mueed15,Mueed15b,Jo17},
in AlAs quantum wells \cite{Shayegan06,Bishop07,Padmanabhan09,Gokmen10}, and, more recently,
in ZnO quantum wells \cite{Maryenko17} have shown that the fractional quantum Hall and CFFS states survive at least up to $\kappa=4-8$, where LL mixing is expected to be substantial. [As discussed in Ref.~\cite{Sodemann13},
the parameter $\kappa$ is given by $2.6/\sqrt{B}$, $14.6/\sqrt{B}$, $16.7/\sqrt{B}$,
and $22.5/\sqrt{B}$ in n-doped GaAs, p-doped GaAs,
n-doped ZnO,and n-doped AlAs, with $B$ measured in Tesla.]
As we will see below, at these $\kappa$ values, LL mixing causes a reduction in the energy of the CFFS by 8-18\%.

\begin{table}[t]
\begin{tabular}{|c|c|c|c|}
\hline
N& $V_C^{\rm proj}(e^2/\epsilon l)$ &  $V_C^{\rm unproj}(e^2/\epsilon l)$ & $\varepsilon_K^{\rm unproj}(\hbar \omega_c)$ \\ \hline
9 & $-0.4705 \pm 0.0002$ & $-0.5067\pm 0.0001$ & $0.089\pm 0.001$ \\ \hline
13 & $-0.4628 \pm 0.0001$ & $-0.5034\pm 0.0001$ & $0.124\pm 0.004$ \\ \hline
25&$ -0.4624 \pm 0.0002$&$-0.5028 \pm 0.0001$&$0.121\pm 0.001$\\ \hline
37&$-0.4659 \pm 0.0002$&$-0.5037\pm 0.0001$&$0.103\pm 0.002$\\ \hline
$\infty$ &  $-0.4657$ & $-0.5034$ & 0.103 \\ \hline
\end{tabular}
\caption{\label{energies} $V_C^{\rm proj}$ and  $V_C^{\rm unproj}$ are the Coulomb interaction energies per particle,
in units of $e^{2}/\epsilon l$, for the projected and the unprojected CF Fermi seas; these include interaction with the uniform positively charged background.
The symbol $\varepsilon_K^{\rm unproj}$ is the kinetic energy per particle for the unprojected CFFS in units of the cyclotron energy $\hbar \omega_c$, measured relative to the LLL.
The numbers in the first three rows are from our calculations in the torus geometry,
whereas the last row shows the thermodynamic limits obtained previously from calculations in the spherical geometry~\cite{Trivedi91,Kamilla97b,Balram17}.
All ground states are in momentum sector (0,0).}
\end{table}

A realistic treatment of LL mixing of the CFFS is a nontrivial task.
Inclusion of higher LLs in exact diagonalization studies severely limits the system sizes that can be studied,
as it leads to an exponential increase in the dimension of the Hilbert space that is to be diagonalized Refs.~\cite{Wojs10,Peterson14}.
A perturbative approach has been developed that simulates weak LL mixing through modification of the interelectron interaction to include three body terms~\cite{Bishara09,Sodemann13,Simon13,Peterson13}.
A non-perturbative approach for treating LL mixing is the so-called fixed phase diffusion Monte Carlo method\cite{Ortiz93,Melik-Alaverdian97,Zhang16,Zhao18}),
which aims to obtain the lowest energy state within the phase sector of a given trial wave function.
The fixed phase method is not appropriate for our purposes,
because we are interested in effects that are dependent on the change in the phase structure of the wave function. 

We consider here a different model for LL mixing that allows a phase variation.
In this model we determine the lowest energy state as a function of $\kappa$ by diagonalizing the Hamiltonian in the subspace defined by two linearly independent states: the LLL projected CFFS and the unprojected CFFS.
In other words,
we consider
\be
\Psi_\beta=\beta \Psi_{\rm{proj}}+(1-\beta) \Psi_{\rm{unproj}},
\ee
and determine the mixing parameter $\beta$ that gives the lowest total energy as a function of $\kappa$. 
(Melik-Alaverdian and Bonesteel\cite{Melik-Alaverdian98} had used a similar model to study the effect of LL mixing for the quasiparticle of the 1/3 state.) This is not an unreasonable model.
The unprojected CFFS wave function is a reasonably good approximation to the CFFS.
Its pair correlation function displays Friedel oscillations with the expected $\pi/k_F$ period~\cite{Kamilla97,Balram15b}
and it has a lower interaction energy than the projected wave function with only a modest amount of kinetic energy cost,
as seen in Table ~\ref{energies}.
(The unprojected wave function actually has lower total energy than the projected wave function for $\kappa> 2.7$.) A hybridization with the unprojected state therefore appears to be a favorable way for the system to lower its energy for finite $\kappa$.
We shall see that this model actually produces lower energies than the fixed phase diffusion Monte Carlo method for a range of parameters.

\begin{figure}[t]
\begin{center}
\includegraphics[width=0.8\columnwidth]{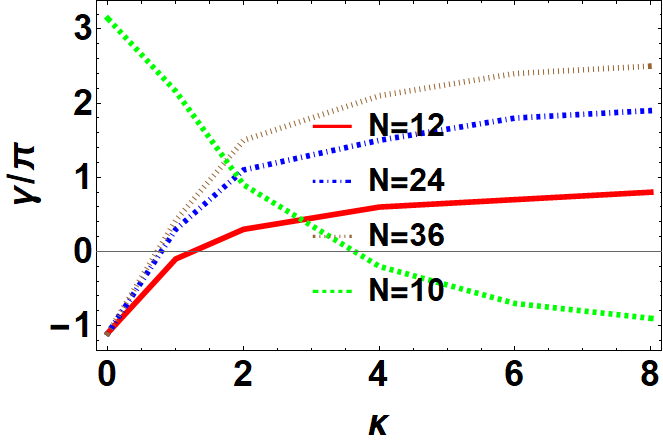} 
\end{center}
\caption{The Berry phase $\gamma$ as a function of the LL mixing parameter $\kappa$  for a Fermi sea with $N=12$,
$N=24$, $N=36$ and $N=10$ composite fermions.
The first three correspond to a closed loop of a CF hole,
while the last one is for the closed loop of a CF particle.
}
\label{partially projected}
\end{figure}

With the wave function so determined, we evaluate the Berry phase corresponding to a closed loop around the Fermi circle following a minor generalization of the prescription of Wang {\em et al.}\cite{Wang17c} and Geraedts {\em et al.}\cite{Geraedts17c}.
The result is shown in Fig.~\ref{partially projected} (details of the calculation are presented later).
A striking feature is the sensitivity of the Berry phase $\gamma$ to the LL mixing parameter $\kappa$.
The variation in the $\gamma$ as a function of $\kappa$ becomes more rapid with increasing $N$.
(Note that while the Berry phases for the projected and the unprojected CFFS wave functions are individually defined only modulo $2\pi$,
the change in the Berry phase during the process of LLL projection can be fully determined by monitoring the Berry phase as a continuous function of the mixing parameter $\beta$ or $\kappa$.) Our study demonstrates an intimate connection of the $\pi$ Berry phase to exact PH symmetry in the LLL.

Two caveats are in order regarding the conclusions in this work. First,
our treatment of LL mixing is, of necessity, approximate,
and it would be important to address the issue by other approaches for treating the effect of LL mixing. Second,
our conclusions are based on a specific definition of the Berry phase for a discrete CFFS,
namely the one used in RefS.~\cite{Wang17c,Geraedts17c}.
It is an interesting question whether an alternative definition would produce a Berry phase that would display a weaker dependence  on $\kappa$.
Ideal would be the calculation of an observable that is manifestly related to the Berry phase. 

For small LL mixings a perturbative approach has been developed which produces a single LL theory but with three and higher body interactions that incorporate the effect of LL mixing and cause a breaking of the PH symmetry.
For completeness,
we have also studied how the Berry phase is modified by the addition of the simplest three-body interaction term that breaks PH symmetry within the LLL. 

The plan of the paper is as follows. In Sec. ~\ref{review} we briefly review the modified LLL projection developed in Ref.~\cite{Pu17}. Following this method,
we construct the CFFS wave function in LLL in Sec. ~\ref{wave function} and show its accuracy by comparing to exact diagonalization. In Sec. ~\ref{LL mixing},
we introduce a treatment of LL mixing by taking superposition of the projected wave functions and unprojected wave functions.
In Sec.~\ref{berry phase} we evaluate the Berry phase as a function of the LL mixing parameter $\kappa$ for several closed paths encircling CFFS.
In Sec.~\ref{PH-breaking}, we ask how the Berry phase is modified by the addition of a three-body interaction that explicitly breaks PH symmetry within the LLL.
Section~\ref{conclusion} concludes the paper with a discussion of the implications of our results,
and also certain caveats.

\section{Brief review of LLL projection in the torus geometry}
\label{review}

In a previous work, PWJ constructed explicit wave functions in the torus geometry for a large class of fractional quantum Hall states and their low energy excitations~\cite{Pu17}.
We describe this construction briefly before extending it to the CFFS. 

A torus is defined by identifying two edges of the parallelogram $\xi_1=L_1$ and $\xi_2=L_1\tau$,
where $\tau$ is a complex number that specifies the geometry of the torus~\cite{Yoshioka83}.
We will work in the symmetric gauge $\vec{A}=(B/2)(y,-x,0)$, which corresponds to a magnetic field $\vec{B}=-B\hat{z}$.
The magnitude $B$ must be chosen so that an integer number $N_\phi=L^2 \rm{Im}(\tau)B/\phi_0$ of flux quanta pass through the system,
with a single flux quantum defined as $\phi_0=hc/e$. 
The single-particle wave functions are chosen to satisfy the boundary conditions \cite{Haldane85,Greiter16} :
\be 
t(L_1)\psi(z,\bar{z})=e^{i\phi_1}\psi(z,\bar{z}), \; t(L_1\tau)\psi(z,\bar{z})=e^{i\phi_\tau}\psi(z,\bar{z})
\ee
where $z=x+iy$ denoted the position of an electron, the phases $\phi_1$ and $\phi_\tau$ define the quasiperiodic boundary conditions,
and $t(L_1)$ and $t(L_1\tau)$ are the magnetic translation operators \cite{Zak64,Brown64},
defined as:
\beq
\label{magnetic translation operator}
t(\xi)&=&\exp^{-\frac{i}{2l^2}\hat{\vec{z}}\cdot(\vec{\xi}\times\vec{r})}T({\xi})\nonumber \\
&=&\exp^{-\frac{i}{2l^2}\hat{\vec{z}}\cdot(\vec{\xi}\times\vec{r})}\exp\left(\xi \partial_z+\bar{\xi} \partial_{\bar{z}}\right)
\eeq
for translation by vector $\vec{\xi}=({\rm Re}(\xi),{\rm Im}(\xi))$.
$T(\xi)$ is the usual translation operator.
A non-trivial aspect of the construction of an N-particle wave function is to ensure that it satisfies the boundary conditions ($j=1,2,\cdots N$):
\beq
&&t_j(L_1)\Psi[z_i,\bar{z}_i]=e^{i\phi_1}\Psi[z_i,\bar{z}_i]\nonumber \\
&&t_j(L_1\tau)\Psi[z_i,\bar{z}_i]=e^{i\phi_\tau}\Psi[z_i,\bar{z}_i]
\eeq

We note that in our convention the magnetic field points in the negative $\hat{z}$ direction.
(The wave functions for a magnetic field pointing in the $+\hat{z}$ direction can be obtained by complex conjugation.) The term  ``counterclockwise" below will refer to counterclockwise rotation relative to the direction of the magnetic field.

We write the single particle states in the symmetric gauge as~\cite{Greiter16}
\be
\psi_i(z,\bar{z})=e^{\frac{z^2-|{z}|^2}{4l^2}}f_i(z,\bar{z}), 
\ee
where the subscript $i$ denotes collectively the LL index and the momentum quantum number.
The explicit form for $f_i(z)$ \cite{Pu17} will not be needed below and so is omitted.
We further denote the wave functions of $n$ filled LLs as 
\be
\Psi_n\equiv \det{\psi_i(z_j,\bar{z}_j)}=e^{\sum_i\frac{z_i^2-|{z}_i|^2}{4l^2}}\chi_n(f_i(z_j)),
\ee
where $\chi_n(f_i(z_j))$ is a Slater determinant formed from $f_i(z_j)$.
The standard unprojected Jain wave functions at $\nu=n/(2pn\pm 1)$ are then constructed as \cite{Jain89,Jain89b,Jain07}
\be
\Psi^{\rm unproj}_{n\over 2pn+1}=\Psi_n \Psi_1^{2p}.
\ee
Here $\Psi_n$ is wave function of $n$ filled LLs in an effective magnetic field corresponding to magnetic flux
\be
\label{number of flux}
N^*_\phi=N_\phi-2p N,
\ee
where $N_\phi$ is the physical magnetic flux quanta number,
and $\Psi_1$ is constructed at magnetic flux $N^{(\nu=1)}_\phi=N$. 
It can be shown that $\Psi^{\rm unproj}$ satisfies the correct boundary conditions \cite{Hermanns13,Pu17}. 

The next step is to project these wave functions into the LLL.
One way is to carry out the so-called ``direct projection"~\cite{Dev92,Wu93,Hermanns13} in which one expands the unprojected wave function in Slater determinant basis functions and retains only the part that resides fully within the LLL.
This method is guaranteed to produce LLL wave function which satisfies the correct boundary conditions. However,
it limits one to very small systems because it requires keeping track of all individual Slater determinant basis functions,
the number of which grows exponentially with system size.
We therefore appeal to the so-called Jain-Kamilla (JK) projection~\cite{Jain97,Jain97b},
which can be implemented for very large systems.
To this end,
we first write the wave function of one filled LL in the Jastrow form:
\be
\Psi_1[z_i,\bar{z}_i]={\cal N} e^{\sum_i\frac{z_i^2-|z_i|^2}{4l^2}}R_1(Z)\prod_{j<k}\theta \left(\frac{z_j-z_k}{L_1}|\tau \right),
\label{LLLstate}
\ee
\beq
\label{F(Z) theta}
R_1(Z)=e^{i{\phi_1-\pi N \over L_1}Z}\theta \left({Z\over L_1}-{\phi_\tau-\phi_1 \tau+\pi N (\tau-1) \over 2\pi}|\tau\right),\nonumber\\
\eeq
where $Z=\sum_{i=1}^N z_i$
is the center of mass coordinate, and $\theta$ is the odd Jacobi theta function\cite{Mumford07}
\be
\theta(z|\tau)=\sum_{n=-\infty}^{\infty}e^{i\pi \left(n+\frac{1}{2}\right)^2\tau}e^{i2\pi \left(n+\frac{1}{2}\right)\left(z+\frac{1}{2}\right)}.
\ee 
satisfying the properties
$
\theta(z+n|\tau)=(-1)^n\theta(z|\tau)
$ and 
$
\theta(z+m\tau|\tau)=(-1)^me^{-i\pi m(2z+m\tau)}\theta(z|\tau)$ for integer $m$ and $n$.
(Our definition of the Jacobi theta function follows the convention of Mumford \cite{Mumford07},
which is different from that used in many other articles in the field of fractional quantum Hall effect.) The standard JK method then suggests the form
$\Psi^{\rm trial}_{n\over 2pn+1}= e^{\sum_i\frac{z_i^2-|z_i|^2}{4l^2}}R_1^{2p}(Z) \chi_n[\hat{f}_{i}(\partial/\partial z_j,z_j) J^p_j]$ with 
\be
J_j=\prod_{k (k\neq j)}\theta \left(\frac{z_j-z_k}{L_1}|\tau \right)
\ee
The resulting wave function, however,  
does not preserve the periodic boundary conditions and thus takes us out of our original Hilbert space.
PWJ overcome this difficulty by noting that a modified wave function, 
\be
\Psi_{n\over 2pn+1} = e^{\sum_i\frac{z_i^2-\bar{z}_i^2}{4l^2}}R_1^{2p}(Z) \chi_{n}(\hat{G}_{i}(\partial/\partial z_j,z_j) J^p_j)
\ee
satisfies the correct boundary conditions. Here, roughly speaking,
the operators $\hat{G}_{i}(\partial/\partial z_j,z_j)$ are obtained from $\hat{f}_{i}(\partial/\partial z_j,z_j)$ by replacing the derivative $\partial/\partial z_j$ by $2\partial/\partial z_j$ whenever it acts on $J^p_j$.
PWJ further demonstrated that the resulting  wave functions for the Jain $n/(2n+1)$ states and their charged and neutral excitations are very accurate representations of the exact Coulomb eigenstates.
The principal advantage of the PWJ construction is that it enables a study of large systems of composite fermions on a torus.
Further details are given in Ref.~\cite{Pu17}.

\section{Construction of wave function for CF Fermi sea}
\label{wave function}

We now construct a LLL wave function for the CFFS. Following the standard method for composite-fermionization,
the unprojected wave function for the CFFS is given by~\cite{Rezayi94}:
\be
\label{unpro}
\Psi_{1\over 2}^{\rm CF}=\det\left[\exp\left(\mi\vec{k}_n \cdot \vec{r}_m\right)\right]\Psi_1^2
\ee
where $\vec{k}$'s can assume values:
\be
\vec{k}=n_1\vec{b}_1+n_2\vec{b}_2 \quad \text{($n_1$ and $n_2$ integers)}
\ee
with
\be
\vec{b}_1=\left({2\pi \over L_1},-{2\pi \mathrm{Re}(\tau) \over L_1\mathrm{Im}(\tau)}\right),\;
\vec{b}_2=\left(0,{2\pi \over L_1\mathrm{Im}(\tau)}\right).
\ee
In the case of rectangular torus in which $\tau$ is purely imaginary, the $\vec{k}$'s are given by:
\be
\left(k_x,k_y\right)=\left({2\pi n_x\over L_1},{2\pi n_y\over L_1|\tau|}\right), \quad \text{($n_1$ and $n_2$ integers)}
\ee
The Fermi sea wave function takes certain values of $\vec{k}$'s to be occupied.
It has been empirically confirmed \cite{Fremling18} that the ground state minimizes $\sum_{i<j}|\vec{k}_i-\vec{k}_j|^2$ as prescribed in Ref.~\cite{Shao15}.
Equation \ref{unpro} is seen to satisfy the correct quasiperiodic boundary conditions because $\det\left[\exp\left(\mi\vec{k}_n \cdot \vec{r}_m\right)\right]$ is purely periodic.

Now we proceed to LLL projection. The wave function consists of terms containing products of $e^{i\vec{k}\cdot\vec{r}}$ and LLL wave functions.
The LLL projection of one such factor $e^{i\vec{k}\cdot\vec{r}} e^{z^2-|z|^2\over 4l^2}f(z)$ produces: 
\be
P_{\rm LLL}e^{i\vec{k}\cdot\vec{r}} e^{z^2-|z|^2\over 4l^2}f(z) =  \hat{F}_k f(z)
\ee
\be
\hat{F}_k = e^{z^2-|z|^2\over 4l^2}e^{-\frac{kl^2}{4}\left(k+2\bar{k}\right)}e^{\frac{\mi}{2}(\bar{k}+k)z}e^{\mi kl^2\partial_z}
\label{eqnhatf}
\ee
where we have used the fact that LLL projection amounts to bringing $\bar{z}$ to the left and making the replacement $\bar{z}\rightarrow 2l^2\partial_z$~\cite{Girvin84b,Jain89}.
With this we can write 
\beq
P_{\rm LLL}\Psi_{1\over 2}^{\rm CF}&=&\det(\hat{F}_{k_n}(z_m))R^2_1(Z)\cdot
\left[\prod_{i<j}\theta\left({z_i-z_j\over L_1}|\tau\right)\right]^2\nonumber \\
&=&\left[R_1\left(Z+\mi l^2\sum_j k_j\right)\right]^2\det(\hat{F}_{k_n}(z_m))\cdot\nonumber \\
&&\left[\prod_{i<j}\theta\left({z_i-z_j\over L_1}|\tau\right)\right]^2 
\label{eq17}
\eeq
So far this equation represents the Direct Projection, which is not amenable to calculations with large systems. 
Following PWJ we make the replacement 
\be
\det(\hat{F}_{k_n}(z_m))\left[\prod_{i<j}\theta\left({z_i-z_j\over L_1}|\tau\right)\right]^2\rightarrow \det(\hat{G}_{k_n}(z_m)J_m)
\ee
where 
\be
\hat{G}_{k_n}(z_m)=e^{z^2-|z|^2\over 4l^2}e^{-\frac{kl^2}{4}\left(k+2\bar{k}\right)}e^{\frac{\mi}{2}(\bar{k}+k)z}e^{\mi 2kl^2\partial_z}
\ee 
is obtained by replacing in Eq.~\ref{eqnhatf} the factor $e^{\mi k_nl^2\partial_{z_m}}$ by $e^{\mi 2k_nl^2\partial_{z_m}}$.
The final expression for the projected CFFS wave function is
\begin{widetext}
\be
\label{CF Fermi sea}
P_{\rm LLL}\Psi_{1\over 2}^{\rm CF}=e^{\sum_i \frac{z_i^2-|z_i|^2}{4l^2}}\left[R_1(Z+\mi l^2\sum_j k_j)\right]^2 \det\left(G_{k_n}(z_m)\right)\\
\ee
\beq
G_{k_n}(z_m) &=& e^{-\frac{k_nl^2}{4}\left(k_n+2\bar{k}_n\right)} e^{\frac{\mi}{2}(\bar{k}_n+k_n)z_m}\cdot 
\prod_{j,j\neq m} \theta \left(\frac{z_m+\mi 2k_nl^2-z_j}{L}|\tau\right)
\eeq
\end{widetext}
Appendix \ref{proof for satisfaction of periodic boundary conditions} contains the proof that this wave function satisfies the correct boundary conditions.
We note that the implementation of JK projection in  Refs.~\cite{Shao15,Geraedts17c,Wang17c} also makes the {\em ad hoc} replacement of $k_n$ by $2k_n$, and,
in that sense, is similar to our wave function.
We have not found a proof that our CFFS wave function is the same as that in Refs.~\cite{Shao15,Geraedts17c,Wang17c},
although numerical comparisons suggest that to be the case at least for the Fermi sea ground state.

Because we have chosen the real axis as our principal axis, the center of mass momentum sectors $(K_1,K_2)$ are characterized by the eigenvalues of $t_{\rm{CM}}\left(L/N_\phi\right)$ and $t_{\rm{CM}}\left(L\tau/N\right)$\cite{Haldane85b,Bernevig12}:
\beq
\label{KX}
t_{\rm{CM}}\left({L\over N_\phi}\right)|\Psi(\vec{K})\rangle&=&\prod_{i=1}^N t_i\left({L\over N_\phi}\right)|\Psi(\vec{K})\rangle \nonumber \\
&=&e^{\mi 2\pi {K_1\over N_\phi}}|\Psi(\vec{K})\rangle
\eeq
\beq
\label{KY}
t_{\rm{CM}}\left({L\tau\over N}\right)|\Psi(\vec{K})\rangle&=&\prod_{i=1}^N t_i\left({L\tau\over N}\right)|\Psi(\vec{K})\rangle \nonumber \\
&=&e^{\mi 2\pi {K_2\over N}}|\Psi(\vec{K})\rangle
\eeq
It can be shown by explicit calculation that $\Psi_{1\over 2}^{\rm CF}$ satisfies Eq.~(\ref{KY}) with $K_2=\sum_n n_2$. On the other hand,
$\Psi_{1\over 2}^{\rm CF}$ does not satisfy Eq.~(\ref{KX}).
Rather,
it only satisfies:
\beq
t_{\rm{CM}}\left(L/N\right)\Psi_{1\over 2}^{\rm CF}=e^{\mi 2\pi {\sum_n n_1\over N}}\Psi_{1\over 2}^{\rm CF}
\eeq
Therefore, it is actually a superposition of CM momentum eigenstate $K_1=\sum_n n_1$ and $K_1=\sum_n n_1+N$.
Nevertheless, we can project $\Psi_{1\over 2}^{\rm CF}$ to the CM momentum eigenstate $K_1=\sum_n n_1$ with the projection operator \cite{Pu17}:
\be
P_1=1+e^{-\mi{\pi \sum_n n_1 \over N}}t_{\rm{CM}}\left({L_1\over N_\phi}\right)
\ee
and to the CM momentum eigenstate $K_1=\sum_n n_1+N$ with the projection operator:
\be
P_2=1-e^{-\mi{\pi \sum_n n_1 \over N}}t_{\rm{CM}}\left({L_1\over N_\phi}\right)
\ee

\begin{figure}[!h]
\begin{center}
\includegraphics[width=7.5CM,height=5.0CM]{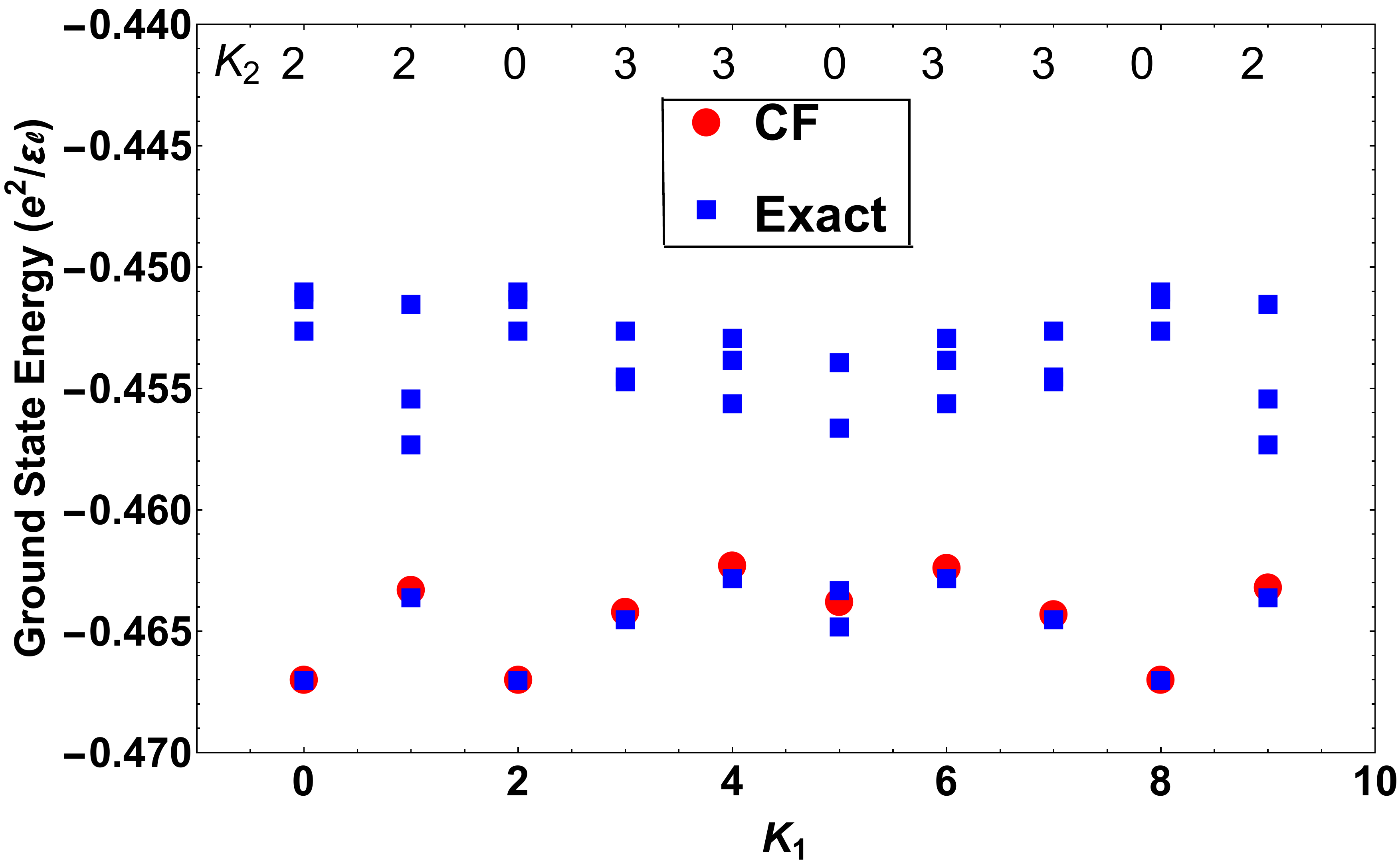} 
\end{center}
\caption{Exact (blue squares) and CF (red dots) energy spectra in momentum sectors $K_1$-$K_2$ for $N=10$ particles exposed to a flux $N_\phi=20$. The momentum $K_1$ is given on the x-axis, and $K_2$ is displayed at the top; for each $K_1$, $K_2$ is chosen to match the momentum of the lowest energy state.  For torus geometry, the spectra for $K_1$ and $K_1+N$ are identical. 
}
\label{ground state energy}
\end{figure}

To test the accuracy of our CF wave function we have compared it with exact Coulomb eigenstates known for small systems.
The comparison of ground-state energies for each momentum sector $K_1$ is shown in Fig.~\ref{ground state energy} for $N=10$ particles.
The wave function clearly is very accurate.

\section{A variational treatment of LL mixing}
\label{LL mixing}

As motivated in the introduction, we define a variational wave function
\be
|\Psi_\beta(\vec{K})\rangle=\beta |\Psi_{\rm{proj}}(\vec{K})\rangle+(1-\beta) e^{-\mi \theta_K}|\Psi_{\rm{unproj}}(\vec{K})\rangle,
\label{mixed}
\ee
where $|\Psi_{\rm{proj}}(\vec{K})\rangle$ is the normalized LLL projected CF wave function and $|\Psi_{\rm{unproj}}(\vec{K})\rangle$ is the normalized unprojected CF wave function.
The phase $\theta_K=\rm{Arg}\langle \Psi_{proj}(\vec{K})|\Psi_{unproj}(\vec{K})\rangle$ 
is introduced to ensure that the phase of the LLL part of the unprojected wave function is the same as that of the projected wave function,
i.e., 
\be
\alpha\equiv \langle \Psi_{\rm{proj}}(\vec{K})|e^{-\mi \theta_K}\Psi_{\rm{unproj}}(\vec{K})\rangle
\ee 
is a positive real number. With this definition, a gauge change for either the projected or the unprojected wave function will leave the Berry phase invariant.

The value of the real parameter $\beta$ is determined by minimization of the total energy.
The expectation value of the kinetic energy per electron (in units of $\hbar \omega_c$),
as measured relative to the lowest LL, is given by 
\be
\varepsilon_K={(1-\beta)^2 \over (1-\beta)^2+\beta^2+2\alpha\beta(1-\beta)} \varepsilon_K^{\rm unproj}
\ee
where $\varepsilon_K^{\rm unproj}=\langle \Psi_{\rm unproj}| H_{\rm kinetic}-1/2|\Psi_{\rm unproj}\rangle$ is the kinetic energy per electron for the unprojected wave function ($\beta=0$). 
Similarly, the interaction energy is given by (in units of $e^2/\epsilon l$)
\be
\varepsilon_V={\beta^2\varepsilon_V^{\rm proj}+(1-\beta)^2 \varepsilon_V^{\rm unproj}+2\beta(1-\beta)\varepsilon_V^{\rm mix}\over (1-\beta)^2+\beta^2+2\alpha\beta(1-\beta)} 
\ee
where $\varepsilon_V^{\rm proj}=\langle\Psi_{\rm{proj}}(\vec{K})|V|\Psi_{\rm{proj}}(\vec{K})\rangle$,
$\varepsilon_V^{\rm unproj}=\langle\Psi_{\rm{unproj}}(\vec{K})|V|\Psi_{\rm{unproj}}(\vec{K})\rangle$,
and $\varepsilon_V^{\rm mix}=\rm{Re}\left[e^{-\mi \theta_K}\langle\Psi_{\rm{proj}}(\vec{K})|V|\Psi_{\rm{unproj}}(\vec{K})\rangle\right]$.
From these, we can obtain the value of optimal $\beta$ as a function of $\kappa$ by minimizing $\varepsilon_K+\kappa \varepsilon_V$.

\begin{figure*}[!h]
\begin{center}
\includegraphics[width=1.5\columnwidth]{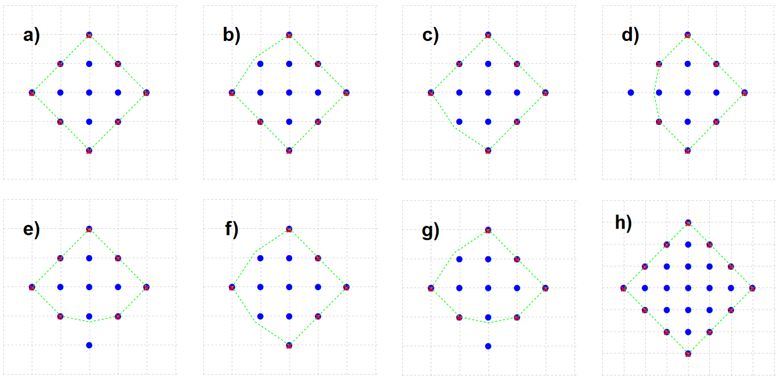} 
\includegraphics[width=1.5\columnwidth]{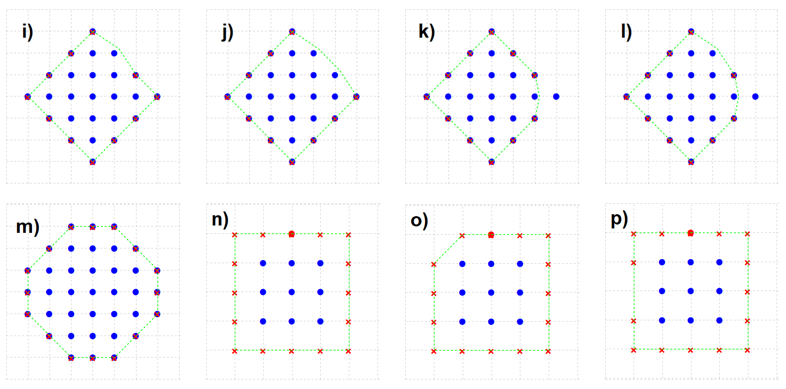} 
\end{center}
\caption{This figure depicts the various paths we have considered for four Fermi sea configurations.
The CF Fermi sea configurations in (a)-(g) consist of $N=12$ composite fermions, those in (h)-(l) have $N=24$ composite fermions, and (m) contains $N=36$ composite fermions; all of these configurations are obtained by creating a CF hole at the Fermi energy of a ``symmetric" CF Fermi sea configuration. 
In all cases, the quoted value of $N$ excludes the CF hole.
 The panels (n)-(p) to a CF Fermi sea  state with $N=10$ composite fermions, including a CF particle (red dot) outside the Fermi contour.
The red crosses connected by dashed green lines indicate the successive positions of the CF hole (particle) along the closed path.}
\label{fig:sea}
\end{figure*}

\begin{figure}[!h]
\begin{center}
\includegraphics[width=0.8\columnwidth]{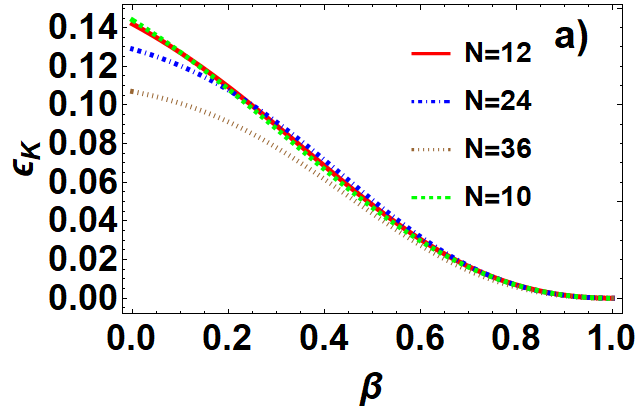} 
\includegraphics[width=0.8\columnwidth]{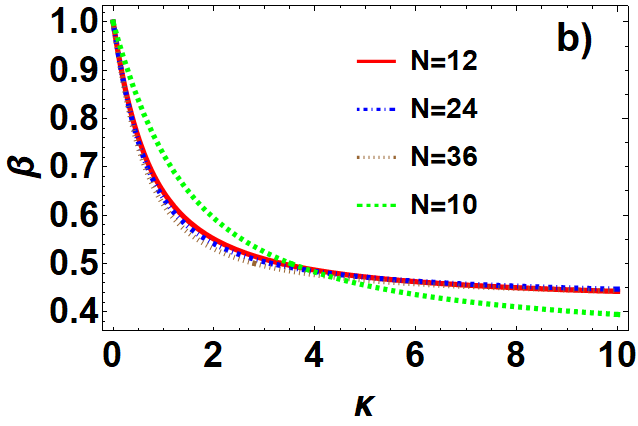} 
\includegraphics[width=0.8\columnwidth]{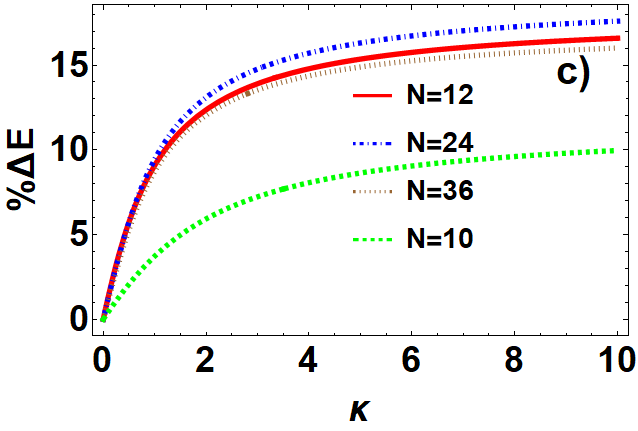} 
\end{center}
\caption{(a) The dependence of kinetic energy per particle $\epsilon_K$ as a function of $\beta$ for $N=12$,
$N=24$,
$N=36$ and $N=10$.
  At $\beta=1$ the CFFS fully projected into LLL, so the kinetic energy is zero,
while at $\beta=0$ we have $\epsilon_K\sim0.1\hbar \omega_c$.
  (b) The dependence of $\beta$ as a function of LL mixing parameter $\kappa$  for $N=12$,
$N=24$,
$N=36$ and $N=10$.
  (c) The \% gain in energy, $\%\Delta E$, as a function of LL mixing parameter $\kappa$ for $N=12$, $N=24$,
$N=36$ and $N=10$.
The energy gain by LL mixing is on the same order as or larger than that found by an earlier diffusion Monte Carlo study\cite{Zhang16}. 
}
\label{parameter}
\end{figure}

We have considered four Fermi sea configurations shown in Figure ~\ref{fig:sea}, with $N=10$, 12,
24 and 36 composite fermions.
(Note that these contain either a CF hole or a CF particle at the Fermi energy.)
The values of $\varepsilon_K^{\rm unproj}$, $\varepsilon_V^{\rm proj}$, $\varepsilon_V^{\rm unproj}$,
$\varepsilon_V^{\rm mix}$ and $\alpha$ are given in Table \ref{parameters}.
Fig.~\ref{parameter}(a) gives the kinetic energy per particle as a function of $\beta$,
and Fig.~\ref{parameter}(b) displays the variational parameter $\beta$ that minimizes the total energy as a function of the LL mixing parameter $\kappa$.
To test the effectiveness of our method in treating LL mixing, 
Fig.~\ref{parameter}(c) shows the percentage reduction in the energy relative to the energy of the fully LLL projected wave function.
The energy reduction is significantly larger than that found by the fixed phase diffusion Monte method,
shown in Fig.~S9 of the Supplemental Material of Ref.~\cite{Zhang16}.

\begin{table}[t]
\begin{tabular}{|c|c|c|c|c|}
\hline
& $N=12$ &  $N=24$&$N=36$&$N=10$\\ \hline
$\varepsilon_K^{\rm unproj}$ & $0.142$ & $0.129$&$0.107$&$0.144$  \\ \hline
$\varepsilon_V^{\rm proj}$&$-0.4597$&$-0.4612$&$-0.4653$&$-0.4623$\\ \hline
$\varepsilon_V^{\rm unproj}$&$-0.5022$&$-0.5023$&$-0.5034$&$-0.5032$\\ \hline
$\varepsilon_V^{\rm mix}$ &  $-0.314$ & $-0.212$&$-0.178$&$-0.303$\\ \hline
$\alpha$&$0.473$&$0.27$&$0.22$&$0.537$\\ \hline
\end{tabular}
\caption{\label{parameters} The values of various quantities needed to determine the lowest energy states as a function of $\kappa$ for the Fermi surface configurations studied in this work.
The energy $\varepsilon_V^{\rm mix}=\rm{Re}\left[e^{-\mi \theta_K}\langle\Psi_{\rm{proj}}(\vec{K})|V|\Psi_{\rm{unproj}}(\vec{K})\rangle\right]$ is quoted in units of $e^2/\epsilon l$. 
The quantities $\varepsilon_V^{\rm proj}=\langle\Psi_{\rm{proj}}(\vec{K})|V|\Psi_{\rm{proj}}(\vec{K})\rangle$,
$\varepsilon_V^{\rm unproj}=\langle\Psi_{\rm{unproj}}(\vec{K})|V|\Psi_{\rm{unproj}}(\vec{K})\rangle$ are given in units of $e^2/\epsilon l$.
We define $\alpha = \langle \Psi_{\rm{proj}}(\vec{K})|e^{-\mi \theta_K}\Psi_{\rm{unproj}}(\vec{K})\rangle$.}
\end{table}

\section{Berry phase in the presence of LL mixing}
\label{berry phase}

We next proceed to evaluate the Berry phase for the wave function defined in Eq.~\ref{mixed}.
With a slight generalization of the convention used by Gearedts {\em et al.}~\cite{Geraedts17c} and Wang {\em et al.}\cite{Wang17c},
we define the Berry phase as:
\be
\label{Berry phase}
\gamma=\sum_{\vec{K}}\rm{Im}\ln \langle \vec{K}+\delta \vec{K}|\hat{\rho}(\delta \vec{K})|\vec{K}\rangle
\ee
where $|\vec{K}\rangle$ denotes the state with a CF hole (a CF particle),
obtained by removing (adding) a composite fermion from (to) a CFFS,
with $\vec{K}$ being the CM momentum of the state.
$\delta \vec{K}$ is the change of the CM momentum as the CF hole or the CF particle moves on the Fermi surface by a discrete step. 
The insertion of the density operator 
\be
\hat{\rho}(\delta \vec{K})=\sum_i \exp\left(\mi \delta \vec{K}\cdot \vec{r}_i\right)
\ee
was motivated in Refs.~\cite{Geraedts17c,Wang17c} as a way to calculate overlap matrix elements between the periodic parts of the two successive ``Bloch" wave functions.

We note that Refs.~\cite{Geraedts17c,Wang17c} used the LLL projected density operator given by
\ba
&&P_{\rm LLL}\hat{\rho}(\vec{k})\\ \nonumber
&=&\sum_i \exp \left(-{kl^2 \over 4}(k+2 \bar{k})\right)\exp \left(\mi z_i(k+\bar{k})/2\right)T_i(\mi kl^2) \\ \nonumber
&=&e^{-|k|^2l^2/4}\sum_i t_i(\mi kl^2)
\ea
For the projected CFFS, the Berry phase is independent of whether one uses the projected or the unprojected density operator,
because then 
$\langle \vec{K}+\delta \vec{K}|\hat{\rho}(\delta \vec{K})|\vec{K}\rangle$ and $\langle \vec{K}+\delta \vec{K}|P_{\rm LLL}\hat{\rho}(\delta \vec{K})|\vec{K}\rangle$
have the same phase (although their moduli are in general different).
The definition of the Berry phase with the unprojected density operator is natural for situations when the CFFS is not confined to the LLL.

The evolution of the Berry phases as a function of $\kappa$ is shown in Fig.~\ref{partially projected} for four CFFSs with  for $N=12$, $N=24$,
$N=36$ and $N=10$.
The Berry phase increases with $\kappa$ when moving a CF hole around,
while decreases with $\kappa$ when moving a CF particle around.
The magnitude of change is seen to be strongly dependent on LL mixing. Furthermore,
the dependence becomes stronger with increasing the number of steps along the Fermi-surface contour, hence also correlated with the system size $N$.

\begin{figure}[!h]
\begin{center}
\includegraphics[width=0.8\columnwidth]{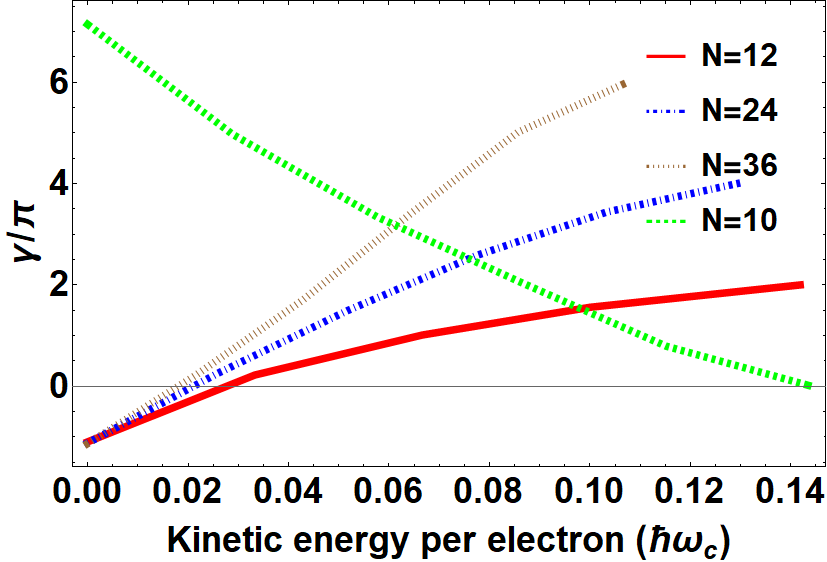} 
\end{center}
\caption{The dependence of Berry phase as a function of kinetic energy per electron  for $N=12$,
$N=24$, $N=36$, and $N=10$ around the paths in Figs.~\ref{fig:sea}(a), \ref{fig:sea}(h), \ref{fig:sea}(m), and \ref{fig:sea}(n).
  The plot for $N=10$ has been translated by $4\pi$ relative to Fig.~\ref{partially projected} to be fitted in the figure.
  The Berry phase rotates by approximately $3\pi$ for $N=12$, $5\pi$ for $N=24$, $7\pi$ for $N=36$,
and $-7\pi$ for $N=10$ as the wave function is continuously projected.
  The Berry phase increases with kinetic energy when moving a CF hole around,
while decreases with kinetic energy when moving a CF particle around.}
\label{kinetic-berry}
\end{figure}

To gain insight into this result, we ask how the Berry phase changes as we vary the wave function continuously from the unprojected to the projected CFFS,
i.e., as the parameter $\beta$ changes from 0 to 1. Figure ~\ref{kinetic-berry} gives the phase variation.
(The Berry phase is shown as a function of the kinetic energy rather than $\beta$,
where the zero kinetic energy point corresponds to $\beta=1$, and the other end corresponds to $\beta=0$.) This suggests that the total change in the Berry phase is given by 
\be
\delta\gamma=\gamma^{\rm unproj}-\gamma^{\rm proj}=\pm\left({N_{\rm steps}\over 2}-1  \right) \pi
\label{deltagamma}
\ee
where $+$ holds for moving a CF hole, $-$ holds for moving a CF particle,
and $N_{\rm steps}$ is the number of discrete steps in the chosen closed path around the Fermi circle.
To test this conjecture, we calculate the Berry phase for a number of other paths,shown in Fig.~\ref{fig:sea}.
(Notice that all our steps are in the counterclockwise direction.)
The results are summarized in Table \ref{step number}.
The fact that $\delta\gamma$ in Eq.~(\ref{deltagamma}) is proportional to $N_{\rm steps}$ explains the rapid variation of the Berry phase as a function of $\kappa$.

The Berry phase for the LLL wave function is consistent with the relation given in Refs.~\cite{Geraedts17c,Wang17c}: 
\be
\gamma^{\rm proj}=\pi\left({N_{\rm{steps}}\over 2}+1\right)\;{\rm mod}\; 2\pi,
\label{gammaproj}
\ee 
The Berry phase for the unprojected wave function is seen to be 
\begin{equation*}
\gamma^{\rm unproj}=\begin{cases}
\pi N_{\rm{steps}}\;{\rm mod}\; 2\pi & \text{if moving a CF hole,}\\
0 & \text{if moving a CF particle.}
\end{cases}
\label{gammaunproj}
\end{equation*}
Both $\gamma^{\rm proj}$ and $\gamma^{\rm unproj}$ are defined only modulo $2\pi$,
in contrast to the difference,
which can be fully determined by tracking the change continuously as a function of $\beta$.

Insight into the Berry phase $\gamma^{\rm unproj}$ for the unprojected wave function can be gained in the following fashion.
The Berry phase for a closed loop of a hole around a (non-relativistic) electron Fermi sea is given precisely by $\gamma^{\text{free electron}}=\pi N_{\rm{steps}}$ mod $ 2\pi$.
It arises from the electron exchange statistics: when a hole moves around a closed loop,
it scrambles the ordering of the electrons along the path,
which produces precisely the phase factor $(-1)^{N_{\rm steps}}$.
Because the additional Jastrow factors in Eq.~\ref{Berry phase} appear as $|{\rm Jastrow}|^2$,
they do not produce any additional phases. (Certain results for an inversion symmetric CFFS are derived in Appendix~\ref{App2}.)
There is thus no ``intrinsic" Berry phase, i.e., the Berry phase apart from the exchange contribution,
for the electron Fermi sea or for the unprojected CFFS.

\begin{table*}
\makebox[\textwidth]{\begin{tabular}{|c|c|c|c|c|c|c|c|c|}
\hline
$N$&path in Fig~\ref{fig:sea} &$N_{\rm{steps}}$&${\gamma^{proj}\over \pi}$ mod 2&$\left({N_{\rm{steps}}\over 2}+1\right)$ mod 2&$ {\gamma^{unproj}\over \pi}$ mod 2 &$N_{\rm{steps}}$ mod 2&${\delta \gamma\over \pi}$&${N_{\rm{steps}}\over 2}-1$\\ \hline
\multirow{4}{*}{12}&a&8 & 0.9&1 & 0&0 & 3.1&3 \\ \cline{2-9}
&b,c,d,e&7& 0.4 &0.5&1&1&2.6&2.5 \\ \cline{2-9}
&f&6&0&0&0&0&2&2\\ \cline{2-9}
&g&6&$-0.1$&0&0&0&2.1&2\\ \hline
\multirow{3}{*}{24}&h&12&$-0.9$&1&0&0&5.1&5\\ \cline{2-9}
&i,k&11&0.4&0.5&1&1&4.6&4.5\\ \cline{2-9}
&j,l&10&$-0.1$&0&0&0&4.1&4\\ \hline
36&m&16&0.9&1&0&0&7.1&7\\ \hline
\multirow{2}{*}{10}&n&16&1.1&1&0&0&$-7.1$&7\\ \cline{2-9}
&o,p&15&0.6&0.5&0&1&$-6.6$&6.5\\ \hline
\end{tabular}}
\caption{\label{step number}The Berry phases for projected and unprojected wave functions,
$\gamma^{\rm proj}$ and $\gamma^{\rm unproj}$,
for the paths shown in Fig.~\ref{fig:sea}.
  These Berry phases are defined modulo $2\pi$. In contrast, the difference $\delta \gamma$ is fully determined by monitoring the phase as the wave function is continuously projected into the LLL.
$N$ is the number of composite fermions and $N_{\rm steps}$ is the number of steps for the closed path in Fig.~\ref{fig:sea}.}
\end{table*}

\begin{figure}[hbt]
\begin{center}
  \includegraphics[width=0.8\columnwidth]{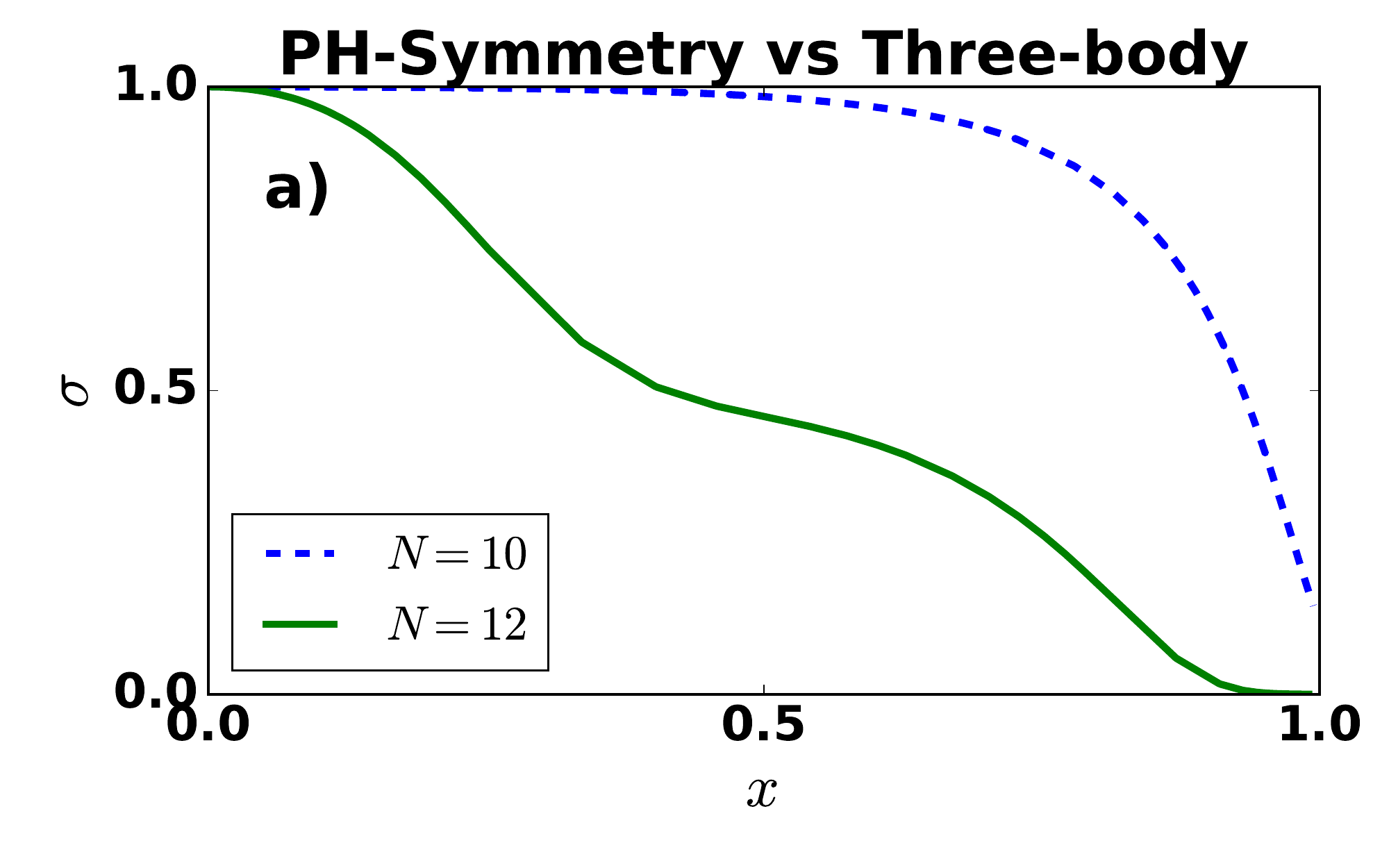}\\
  \includegraphics[width=0.8\columnwidth]{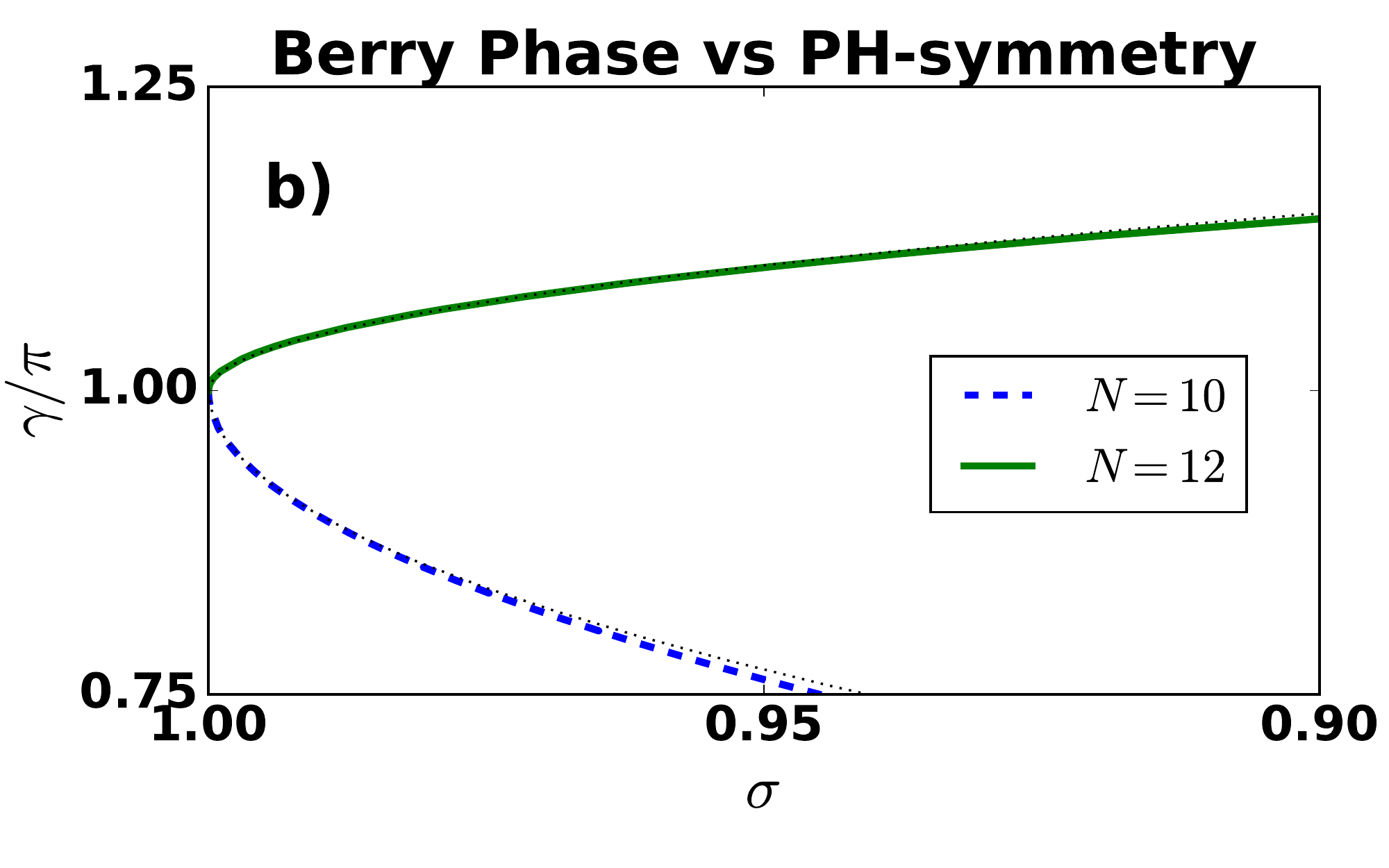}\\
  \includegraphics[width=0.8\columnwidth]{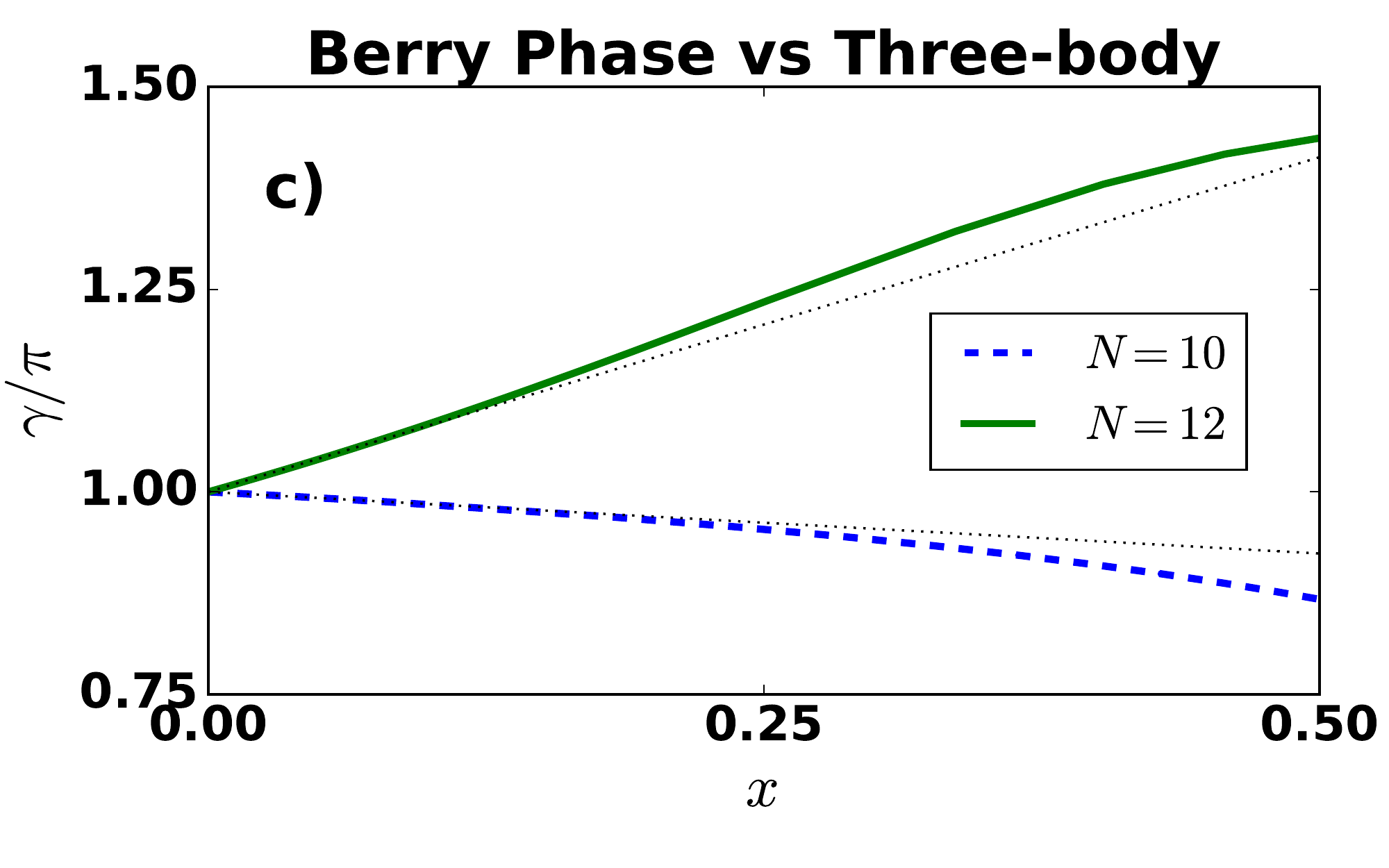} 
\end{center}
\caption{Variation in the Berry Phase $\gamma$ for the CF Fermi sea for the PH-perturbed Hamiltonian $H=(1-x)H_{\mathrm{Coulomb}} + xH_{\rm 3B}$,
  where $x$ sets the strength of the three-body interaction $H_{\rm 3B}$.
  Panel (a) shows the parameter $\sigma$ as a function of $x$, where $\sigma$ is so defined (see text) that its deviation from unity is a measure of the degree of PH-symmetry breaking.
Panels (b) and (c) depict the change in the Bery phase $\gamma$ as a function of $\sigma$ and $x$. 
}
\label{fig:berry_ph}
\end{figure}

\section{Berry Phase under PH-Symmetry-breaking perturbations}
\label{PH-breaking}

In this section we probe the sensitivity of the Berry Phase to PH-symmetry breaking.
We add to the Coulomb potential a weak three-body interaction,
which breaks the PH symmetry of the exact energy eigenstates.
We then use the perturbed ground states to perform the Berry phase calculation. 

We know that, in the absence of LL mixing, the projected CFFS has almost perfect overlap with the exact eigenstates.
This means that we can replace the various states along the path under consideration by the corresponding lowest energy eigenstates \cite{Geraedts17c}.
We may now probe the PH-symmetry dependence of the Berry phase $\gamma$ by weakly perturbing the exact sates away from the PH-symmetric point.

We introduce the Hamiltonian 
\be
H = (1-x)H_{\mathrm{Coulomb}} + xH_{\rm 3B} 
\ee
where $H_{\rm 3B}=V_{\rm 3B} |\Psi^{(3)}_{L=3}\rangle \langle \Psi^{(3)}_{L=3}|$ is the three-body interaction that penalizes the three body wave function $\Psi^{(3)}_{L=3}$ in the angular momentum three channel.
We choose $V_{\rm 3B}$  such that the gap in the ${\vec K}=(N/2,N/2)$ sectors are equal for $H_{\rm 3B}$ and $H_{\mathrm{Coulomb}}$,
if the latter is projected onto LL $n=1$ instead of $n=0$. Note that both $H_{\rm Coulomb}$ and $H_{\rm 3B}$ produce incompressible states in the 2nd LL.
If we send $x\to1$, we will reach the Pfaffian state (or one of its excitations),
but for small $x$ the CFFS state should be only mildly perturbed.

We consider here the paths in Figs.~\ref{fig:sea} (n) and (a) corresponding to $N=10$ and $N=12$ particles. 
We define $\sigma$ as a measure of the PH symmetry:
\be
\sigma=\frac1{N_{\rm K}}\sum_{\vec K}\left|\langle\Psi(-\vec K)|\mathcal C|\Psi(\vec K)\rangle\right|^2
\ee
where $\mathcal C$ performs the PH-conjugation, $\vec K$ are the momenta of the configurations along the path,
and $N_{\rm K}$ is the number of such momenta. 
The value $\sigma=1$ corresponds to perfect PH symmetry.
The relation between $\sigma$ and $x$ is shown in Fig.~\ref{fig:berry_ph}(a) for the two paths depicted in Figs.~\ref{fig:sea}(a) and \ref{fig:sea}(n).
Figure.~\ref{fig:berry_ph}(b) gives the variation of the Berry phase $\gamma$ as a function of $\sigma$,
and Fig.~\ref{fig:berry_ph}(c) as a function of $V_{\rm 3B}$.  
The behavior of $\gamma$ as a function of $\sigma$ is well approximated by 
\be \gamma=\pi-1.03\pi\sqrt{1-\sigma}, \;\; N=10
\ee
\be
\gamma=\pi+0.45\pi\sqrt{1-\sigma},\;\; N=12
\ee
Even though the PH symmetry is being broken in a specific manner in this model,
our calculation demonstrates a connection between $\pi$ Berry phase and exact PH symmetry.
We find that $\gamma$ varies very rapidly with $\sigma$ for small $1-\sigma$.

\section{Conclusions}
\label{conclusion}

We have extended the construction of PWJ~\cite{Pu17} to obtain LLL projected wave functions for the CFFS and its excitations in the torus geometry.
Explicit comparison shows that these are accurate representations of the exact Coulomb eigenstates. 

We have considered a model for LL mixing in which the projected CFFS can hybridize with the unprojected CFFS.
The resulting wave function indicates a substantial lowering of energy at finite $\kappa$.
Within this model, the Berry phase of the CFFS is found to vary rapidly with LL mixing,
illustrating an intimate connection between the $\pi$ Berry phase and PH symmetry in the LLL.
We stress that this conclusion relies on using the prescription of Geraedts {\em et al.}\cite{Geraedts17c} for defining the Berry phase,
and we do not rule out the possibility that an alternative definition of the Berry phase would make it more robust to LL mixing. 

One may ask whether our conclusions would carry over to a more realistic model for incorporating LL mixing.
Another treatment of LL mixing is through the fixed phase diffusion Monte Carlo method,
which, as mentioned in the introduction,
does not allow the phase of the wave function to change and may thus not be appropriate for the question of the Berry phase.
However, because our method produces lower energies than the fixed phase diffusion Monte Carlo method,
we believe it is likely that LL mixing does cause a change in the phase of the CFFS wave function.
With this in mind, our work shows, at minimum,
that there exists an energetically favorable CFFS wave function that,
when hybridized with the LLL projected CFFS,
will produce a rapid variation of the Berry phase with LL mixing.
The sensitivity of Berry phase to LL-mixing is also tested in a second method by adding a weak three-body interaction.
We show the Berry phase varies with the PH breaking degree.

\section{Acknowledgments}

This work was supported in part by the U. S. National Science Foundation under Grant No.
DMR-1401636 and by the Science Foundation Ireland Principal Investigator Award 12/IA/1697.
We thank Ajit Balram, Senthil Todadri, Chong Wang, Jie Wang, Ying-Hai Wu and Jiabin Yu for illuminating discussions, generous help and advice,
and Ajit Balram for providing the thermodynamic energy of the unprojected CFFS.

\appendix

\section{Boundary conditions for the CF Fermi sea wave function in Eq.~\ref{CF Fermi sea}}
\label{proof for satisfaction of periodic boundary conditions}

It is straightforward to confirm the boundary conditions on the real axis:
\be
t_i(L_1)P_{\rm LLL}\Psi_{1\over 2}^{\rm CF}=P_{\rm LLL}\Psi_{1\over 2}^{\rm CF}
\ee
For the other direction we get:
\beq
T_m(L_1\tau)&&G_{k_n}(z_m) = e^{\mi 2\pi n_1\tau}e^{-\mi \pi(\tau+1)(N-1)} \nonumber \\
&&\times \prod_{j,j\neq m}e^{-\mi 2\pi{z_m+\mi 2k_nl^2-z_j \over L_1}}G_{k_n}(z_m)
\eeq
and
\be
T_m(L_1\tau)G_{k_n}(z_r)=e^{-\mi \pi(\tau+1)}e^{i2\pi{z_r+i2k_nl^2-z_m \over L_1}}G_{k_n}(z_r)
\ee
These relationships imply
\beq
\label{trans tau}
T_m(L_1\tau)\det(G_{k_n}(z_m))&=&e^{-\mi 2\pi(\tau+1)(N-1)}e^{{\mi 4\pi\over L_1}\sum_{j}z_j}e^{-\mi {4\pi\over L_1}Nz_m} \nonumber \\
&&e^{{-4\pi l^2\over L_1}\sum_j k_j}\det(G_{k_n}(z_m))
\eeq
On the other hand, periodic boundary condition requires
\beq
&&T_m(L_1\tau)\left[R_1(Z+\mi l^2\sum_j k_j)\right]^2 \det\left(G_{k_n}(z_m)\right)=\nonumber \\
&&e^{-\mi \pi2N({2z_m \over L_1}+\tau)}\left[R_1(Z+\mi l^2\sum_j k_j)\right]^2 \det\left(G_{k_n}(z_m)\right)\nonumber \\
\eeq
With 
\be
T_m(L_1\tau)\left[R_1(Z+\mi l^2\sum_j k_j)\right]^2=e^{-\mi 2\pi(\tau+{2\over L_1}(Z+\mi l^2\sum_j k_j))}
\ee
this is equivalent to 
\beq
\label{pbc tau}
T_m(L_1\tau)\det\left(G_{k_n}(z_m)\right)&=&e^{-\mi 2\pi\tau(N-1)}e^{-\mi{4\pi N z_m\over L_1}}e^{\mi{4\pi\over L_1}\sum_j z_j}\nonumber \\
&&e^{-{4\pi l^2\sum_j k_j\over L_1}}\det\left(G_{k_n}(z_m)\right)
\eeq
Equation ~(\ref{trans tau}) and (\ref{pbc tau}) are identical, proving that the wave function in Eq.~(\ref{CF Fermi sea}) satisfies the correct periodic boundary conditions in the $L_1\tau$ direction as well.

\section{Berry phase for an inversion symmetric system}
\label{App2}
We show below that the overlap $ \langle \vec{K}+\delta \vec{K}|\hat{\rho}(\delta \vec{K})|\vec{K}\rangle$ for each step is real,
on condition that the ground state is inversion symmetric (\eg all paths shown in Fig.~\ref{fig:sea}) and $\phi_1=0$, $\phi_\tau=0$.
We further show that the Berry phase for the unprojected wave function is trivial, \ie $0$ mod $2\pi$,
if the path is also inversion symmetric (\eg (a), (h), (m) and (n) in Fig.~\ref{fig:sea}).

We set the origin of the momentum coordinate system at the the inversion symmetry center of the CF Fermi sea ground state (\ie the state without the additional CF hole or CF particle).
Consider a path in which a CF hole (particle) is moved from $\vec{K}_1$ to $\vec{K}_2$.
These two states are labeled as $|\vec{K}_1\rangle$ and $|\vec{K}_2\rangle$.
The inner product can be written explicitly as:
\beq
&&\langle \vec{K}_2 |\hat{\rho}(\vec{K}_2-\vec{K}_1)| \vec{K}_1 \rangle \\ \nonumber
&=&\int d^2\vec{r_1}\cdots d^2\vec{r_N} \left[\det(\vec{K}_2)\Psi_1^2\right]^*\\ \nonumber
&&\hat{\rho}(\vec{K}_2-\vec{K}_1)\left[\det(\vec{K}_1)\Psi_1^2\right]
\eeq
Here $\det(\vec{K})$ represents the determinant of plane waves corresponding to the occupied momenta of $|\vec{K}\rangle$.
Another segment of the path is the inversion-symmetric partner from 
$\vec{K}'_1=-\vec{K}_1$ to  
$\vec{K}'_2=-\vec{K}_2$. On condition that $\phi_1=0$, $\phi_\tau=0$, $\Psi_1^2$ is even under inversion.
With transformation of integration variables $\vec{r_i}\rightarrow -\vec{r_i}$,
$i=1,2 \cdots N$,
we get:
\be
\label{inv sym1}
\langle \vec{K}_2 |\hat{\rho}(\vec{K}_2-\vec{K}_1)| \vec{K}_1 \rangle = \langle \vec{K}'_2 |\hat{\rho}(\vec{K}'_2-\vec{K}'_1)| \vec{K}'_1 \rangle.
\ee 
On the other hand, for the unprojected wave functions, we have:
\be
\hat{\rho}(\vec{K}_2-\vec{K}_1)=\hat{\rho}(\vec{K}'_2-\vec{K}'_1)^*
\ee
and
\be
\det(\vec{K}'_1)=(-1)^m \det(\vec{K}_1)^*
\ee
\be
\det(\vec{K}'_2)=(-1)^m \det(\vec{K}_2)^*
\ee
where $m$ can be an odd or even integer depending on the ordering of the momenta.
With these results, we get:
\be
\label{inv sym2}
\langle \vec{K}_2 |\hat{\rho}(\vec{K}_2-\vec{K}_1)| \vec{K}_1 \rangle = \langle \vec{K}'_2 |\hat{\rho}(\vec{K}'_2-\vec{K}'_1)| \vec{K}'_1 \rangle^*.
\ee
Eq.~\ref{inv sym1} and Eq.~\ref{inv sym2} together tell us that $\langle \vec{K}_2 |\hat{\rho}(\vec{K}_2-\vec{K}_1)| \vec{K}_1 \rangle=\langle \vec{K}_2 |\hat{\rho}(\vec{K}_2-\vec{K}_1)| \vec{K}_1 \rangle^*$, which means it must be real,
\ie  the phase of overlap can only be $0$ or $\pi$ mod $2\pi$ for unprojected wave functions. Furthermore,
since Eq.~(\ref{inv sym1}) tells us the $\pi$ phase must come in pairs for an inversion symmetric path,
the Berry phase for the unprojected wave function must be $0$ mod $2\pi$ for such paths.


\end{document}